# Interpretable AI-Driven Discovery of Terrain-Precipitation Relationships for Enhanced Climate Insights

Hao Xu[1], Yuntian Chen[2,*], Zhenzhong Zeng[3], Nina Li[4], Jian Li[5,*], and Dongxiao Zhang[2,6,*]

[1] BIC-ESAT, ERE, and SKLTCS, College of Engineering, Peking University; Beijing 100871, P. R. China

[2] Ningbo Institute of Digital Twin, Eastern Institute of Technology; Ningbo, Zhejiang 315200, P. R. China

[3] South University of Science and Technology of China; Shenzhen, Guangdong 518055, P. R. China

[4] National Meteorological Center, China Meteorological Administration; Beijing 100081, China

[5] Chinese Academy of Meteorological Sciences; Beijing 100081, China

[6] Department of Mathematics and Theories, Peng Cheng Laboratory; Shenzhen 518000, Guangdong, P. R. China

* Corresponding authors. Email address: ychen@eitech.ac.cn (Y. Chen); lij@cma.gov.cn (J. Li); dzhang@eitech.edu.cn (D. Zhang).

November 18, 2023

**Abstract:**

Despite the remarkable strides made by AI-driven models in modern precipitation forecasting, these black-box models cannot inherently deepen the comprehension of underlying mechanisms. To address this limitation, we propose an AI-driven knowledge discovery framework known as genetic algorithm-geographic weighted regression (GA-GWR). Our approach seeks to unveil the explicit equations that govern the intricate relationship between precipitation patterns and terrain characteristics in regions marked by complex terrain. Through this AI-driven knowledge discovery, we uncover previously undisclosed explicit equations that shed light on the connection between terrain features and precipitation patterns. These equations demonstrate remarkable accuracy when applied to precipitation data, outperforming conventional empirical models. Notably, our research reveals that the parameters within these equations are dynamic, adapting to evolving climate patterns. Ultimately, the unveiled equations have practical applications, particularly in fine-scale downscaling for precipitation predictions using low-resolution future climate data. This capability offers invaluable insights into the anticipated changes in precipitation patterns across diverse terrains under future climate scenarios, which enhances our ability to address the challenges posed by contemporary climate science.

*Keywords*. Knowledge discovery techniques, climate processes, terrain complexity, precipitation patterns, climate change

**Introduction**

Precipitation is as a pivotal component for characterizing the climate system and its inherent variability, with far-reaching implications for global climate changes(*1*, *2*), water availability(*3*), ecosystems(*4*), and human societies(*5*). Moreover, calamities, such as floods and mudslides usually emerge in complex terrain, which necessitates a deeper terrain-precipitation understanding to diminish their influence. Despite the remarkable strides made by AI-driven models in modern precipitation and global weather forecasting(*6–9*), these black-box models cannot inherently deepen the understanding of natural phenomena and their underlying mechanisms. While predictions obtained on statistical principles are feasible, it is the formulation of physical relationships that truly fosters an enriched understanding of the underlying mechanisms, which constitutes a key step from observed phenomena to essential principles. For example, humans have historically been able to predict imminent precipitation from thunderstorms through experience and statistical principles. However, the fundamental understanding of the connection between thunderstorms and precipitation was achieved much later in history through the formulation of physical equations. Given that the formation of precipitation involves a complex interplay of various elements within the climate system, uncovering an explicit equation to describe the relationship between precipitation and terrain remains a pressing challenge of our time.

Throughout history, human comprehension of precipitation mechanisms has steadily accumulated and expanded (Fig. 1A). Initially, humans discerned certain patterns through extensive observations of precipitation. Subsequently, with advancements in the field of physics, the formulation of the Clausius-Clapeyron equation provided a robust theoretical foundation for developing mathematical models of precipitation(*10*, *11*). However, in the case of an intricate precipitation system, manually derived equations often necessitate specific assumptions or simplified conditions, limiting their explanatory capacity and scalability. Fortunately, the advent of computer technology has unlocked new possibilities for unravelling the intricate relationship between precipitation and the diversity of influencing factors, such as terrain and climate.

Geographic regression models became popular in research regarding precipitation to address the spatially varied land surface characteristics(*12–15*), which can offer explicit expressions that possess clear physical interpretations through multilinear regression(*16*) or geographic weighted regression(*17*). While these models contribute to understanding the laws and mechanisms governing terrain's influence on precipitation, they often rely on physical explanatory terms proposed by human researchers(*18–20*). Moreover, these models often resort to overly simplistic linear formulations, resulting in limited accuracy. Given the intricate relationship between terrain and precipitation, manually selecting the optimal formula from an immense number of potential variable combinations is a formidable challenge. The emergence of artificial intelligence (AI) technology has opened new avenues for accurately predicting precipitation patterns across diverse terrains. Some studies have successfully utilized emerging AI methods to directly forecast precipitation by leveraging physical measurements, such as radar echoes(*8*) and climate data(*21*). These AI-based approaches have made remarkable strides in achieving high-precision precipitation forecasting (Fig. 1A). However, a noteworthy drawback of such methods is their reliance on black-box models, which often lack interpretability and fail to generate new knowledge in mechanisms. Consequently, the interpretability and reliability of the results may be compromised, limiting their broader applicability.

Considering the abovementioned pain points, our study aims to establish an AI knowledge discovery framework to combine the power of AI-driven models with the interpretability and physical understanding of known explicit mathematical expressions. As a relatively recent concept, knowledge discovery leverages the power of machine learning algorithms to unearth hidden physical laws from observation data directly. While previous knowledge discovery studies predominantly concentrated on exploring traditional partial differential equations within physics(*22, 23*), they largely remained in the proof-of-concept stage where the known equations were rediscovered to confirm the effectiveness of the methodologies. Thus, unknown equations have not been discovered to uncover new insights or knowledge. In this work, we extend the scope of knowledge discovery technology to geography to uncover hidden laws and patterns within geographical data. Specifically, we proposed an AI-driven knowledge discovery framework called genetic algorithm-geographic weighted regression (GA-GWR), which unravels explicit equations to describe the underlying laws governing the relationship between precipitation patterns and terrain characteristics for the first time (Fig. 1B). These previously unknown equations harmonize interpretability and accuracy and perform better than empirical formulas in fitting precipitation data. The AI-empowered discovery of terrain-precipitation equations makes two important contributions. It advances the practical application of existing proof-of-concept knowledge discovery algorithms, and this finding takes a step closer to the essence of the terrain-precipitation interaction than prevailing black-box models employed for precipitation prediction (Fig. 1B).

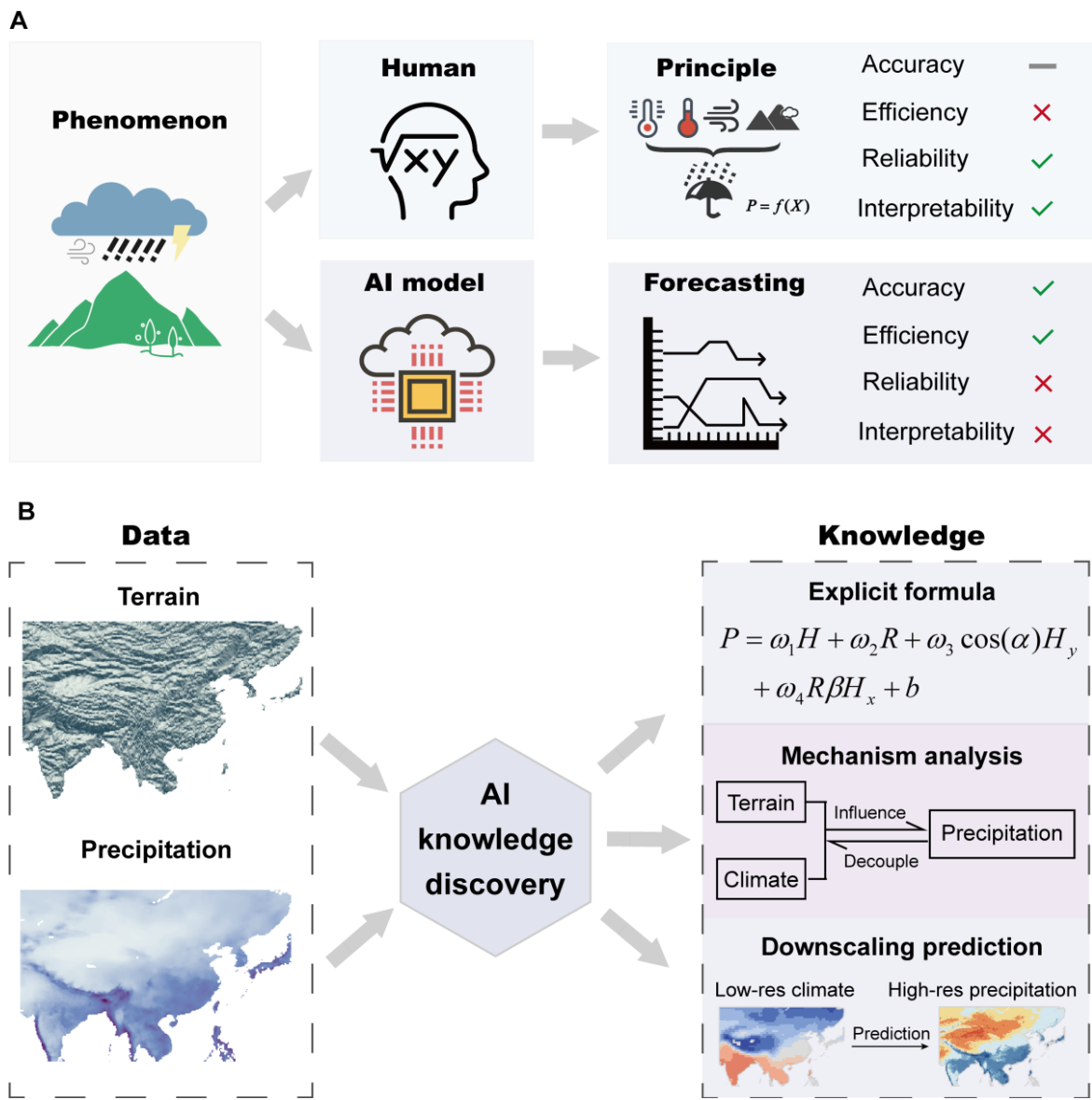

**Fig. 1. Two approaches to understanding the terrain-precipitation relationship and the framework of artificial intelligence (AI) knowledge discovery in this study. (A)** Contrasting the merits and limitations of manual principles summarization against the development of AI

models for predictive analysis. **(B)** Illustrating the framework and key accomplishments facilitated by AI-driven knowledge discovery in this research endeavour.

## Results

**Discovery of new terrain-precipitation formula**

Four fundamental terrain parameters, namely, altitude, relief, slope, and aspect, are employed for high-precision grid-based parameterization. The annual precipitation data are obtained from the Asian Precipitation-Highly Resolved Observational Data Integration Toward Evaluation of Water Resources Project (APHRODITE)(*24*). In this context, the proposed AI-empowered knowledge discovery framework, GA-GWR (for details, refer to *Methods*), is employed to automatically uncover the underlying governing equation directly from the terrain and the average precipitation data from 1951 to 2015 (a 65-year period). As illustrated in Fig. 2A, the terrain variables are digitized as fundamental genes to form different equations (i.e., genomes). The GA-GWR utilizes a genetic algorithm to evolve the optimal equation with the smallest fitness, measured by both fitting precision and parsimony. The GA-GWR procedure merges the flexibility and large search scope of the GA and the adaptability of GWR to spatially varied land surface characteristics. Ultimately, the following terrain-precipitation formula is discovered:

$$P = \omega_1 H + \omega_2 R + \omega_3 \cos(\alpha) H_y + \omega_4 R \beta H_x + b, \qquad (1)$$

where *P*, *H*, *R*, *α*, and *β* are the average annual precipitation, altitude, relief, slope, and aspect, respectively; $H_x$ and $H_y$ are the altitude gradient between grids in the longitude and latitude directions, respectively. The optimization process is summarized in Table S3. In Eq. (1), the coefficients *ω* and *b* are both spatial fields obtained from GWR. From Fig. 2B, the newly discovered formula can fit well with the observation data. The coefficient of determination ($R^2$) of the fitted value is 0.991, while the relative error in most grids is under 5% (Fig. 2C).

Compared to previously proposed empirical formulas(*25, 26*) and baselines, our discovered formula demonstrates superior precision in fitting precipitation data within mesoscale terrains, even utilizing fewer terms (Supplementary Information S1.2). This achievement can be attributed to the effectiveness of our AI-empowered knowledge discovery framework in identifying equations that strike a balance between parsimony and accuracy from the vast number of possible equation combinations. In contrast to traditional methods of manually determining potential terms and conducting linear regression, our approach automatically discovers new yet effective interaction terms (i.e., $\cos(α)H_y$) and utilizes machine learning techniques to automatically generate the optimal formula, thus improving the efficiency and accuracy.

The explicit physical meanings associated with each term in the equation allow for a deeper analysis of the spatial distribution of their coefficients, thereby unveiling the underlying mechanism through which terrain influences precipitation. By scrutinizing the spatial patterns and magnitudes of these coefficients, we gain valuable insights into the intricate processes that govern the interaction between terrain characteristics and precipitation. Because we are examining the average annual precipitation across each grid point over 65 years, Eq. (1) provides a comprehensive depiction of the long-term relationship between terrain and precipitation within the climate system. As depicted in Fig. 2D, the distribution of the constant term *b* aligns with the observed precipitation patterns across diverse climate zones, effectively capturing the terrain-independent component of precipitation. This consistency highlights the robustness of our approach in delineating the contribution of factors beyond terrain characteristics, such as atmospheric dynamics and large-scale climate patterns, to the overall precipitation distribution.

Altitude (*H*) and relief (*R*) play a dominant role in shaping precipitation over complex terrain, especially in northeast and middle east China, the Himalayas, northwest of the Indo-China Peninsula, and the Deccan Plateau in India (Fig. 2D). By using the Himalayas as an example, we can observe notable disparities in precipitation patterns between its southern and northern slopes, which reflects the prominent rain shadow effect. Another noteworthy observation is the inverse relationship between the coefficients of relief and altitude. This observation suggests that relief acts as a compensatory factor, countering the influence of altitude on precipitation to some extent. The last two gradient terms, i.e., $\cos(α)H_y$ and $Rβ H_x$, primarily represent the influence of larger-scale variations in terrain relief and slope on precipitation. As the Asian monsoon system primarily carries moisture from the oceans, these terms capture the components of water vapour interception by intermediate-scale terrain, which in turn affects precipitation patterns. The coefficient of the cos(α)$H_y$ term is more pronounced and exhibits a positive correlation with the overall coefficient $\omega_1$, which reflects that the east–west component of the Asian monsoon is more influential. Consequently, the coefficient $\omega_4$ is much smaller in comparison. Moreover, additional experiments confirmed the significance of both altitude gradient terms in enhancing the descriptive ability of the formula (Supplementary Information S1.4).

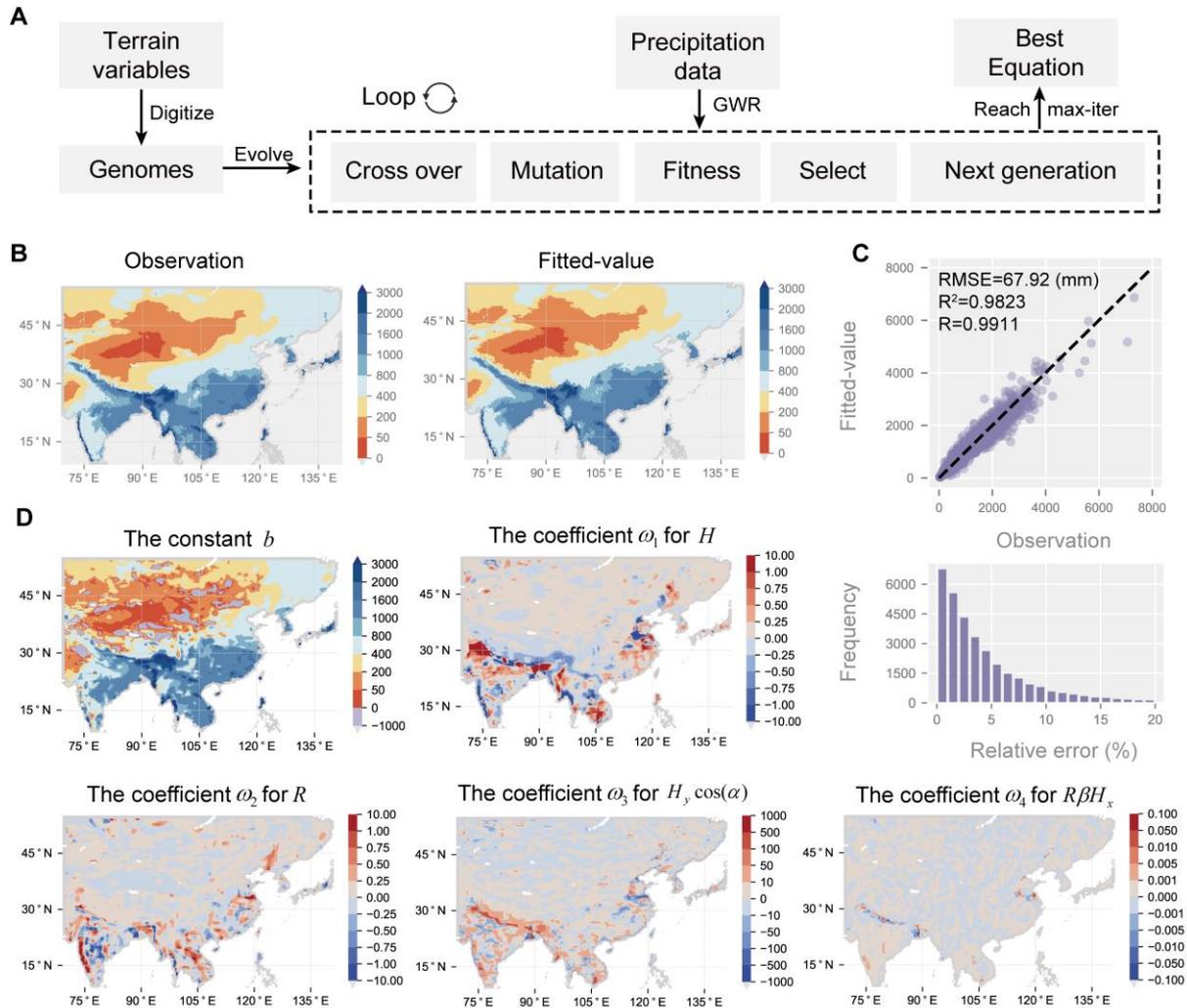

**Fig. 2. The results of artificial intelligence (AI) knowledge discovery. (A)** Flow chart of AI-empowered knowledge discovery. **(B)** The observed and fitted average annual precipitation field from the discovered terrain-precipitation formula. **(C)** The observed and fitted average annual precipitation for each grid data and the distribution of relative error. **(D)** The spatial field of the fitted coefficients $\omega$ and $b$. The grey grids refer to the locations that lack sufficient observation points for regression.

**Influence of climate change on the terrain-precipitation relationship**

Mean annual precipitation exhibits temporal variability due to variations in meteorological conditions over the years. To delve deeper into this complex spatial-temporal relationship between terrain and annual precipitation, we use GA-GWR to automatically seek an optimal equation that captures this evolving spatial-temporal relationship (Fig. 3A). Thus, the fitness function is modified to encompass the mean fitting error of the equation for annual precipitation for each year (for details, refer to *Methods*). This modification ensures that the optimal equation not only demonstrates strong explanatory power for the overall trend but also exhibits good performance in explaining the fluctuations observed in annual precipitation across years. We utilize annual precipitation data spanning from 1951 to 2010, covering 60 years, to uncover the underlying formula. Subsequently, data from the years 2011 to 2015 (5 years) are employed to test the performance and effectiveness of the discovered equation. The resulting spatiotemporal terrain-precipitation formula is expressed as follows:

$$P(t) = \omega_1(t)H + \omega_2(t)R + \omega_3(t)\cos(\alpha)H_y + \omega_4(t)R\beta H_x + b(t), \qquad (2)$$

where $P(t)$ is the annual precipitation varying with year $t$; the coefficients $\omega(t)$ and $b(t)$ are both spatial-temporal fields for year $t$ obtained from GWR. The optimization process is summarized in Table S4.

An intriguing observation emerges, where the structure of the discovered spatial-temporal equation resembles that of Eq. (1), although they are optimized independently. This compelling consistency verifies the capacity and stability of the equation's structure in describing the intricate relationship between terrain and precipitation. Different from Eq. (1), Eq. (2) is a spatial-temporal equation that can describe nonlinear temporal changes in precipitation. The mapping relationship between different physical quantities can be represented by time-varying coefficients without affecting the structure of the equation. This formula learns invariance by incorporating time-varying effects into the coefficients while maintaining the structure of the equation unchanged. From Fig. 3B, the discovered equation demonstrates satisfactory fitting accuracy on both the training and testing datasets, which confirms that it captures the spatial-temporal variations in the terrain-precipitation relationship.

By leveraging Eq. (2), we obtain spatial-temporal coefficients that dynamically change over the years, which enables temporal variation analysis for these coefficients. Here, we primarily focus on studying the temporal variation in the coefficient $\omega_1(t)$, which represents the influence of altitude on precipitation. To better demonstrate its changing dynamics, we adopt the coefficients of normalized terms and take a moving 30-year average in the region to study the impact of climate change. Fig. 3D illustrates the variation in the average normalized coefficient $\omega_1(t)$ with the end year of the time range for the entire Asian region. Notably, we observe a significant downwards trend, indicating a gradual weakening of the overall impact of altitude on precipitation. However, a particularly interesting finding emerged in approximately 1995, which we refer to as the "1995 turning point." After this point, different terrain regions (Fig. 3C) experienced distinct changes in the relationship between altitude and precipitation (Fig. 3E). For example, in the Changbai Mountains (Region 1) and the Himalayas (Region 5), the impact of local altitude on precipitation shows a trend of first decreasing and then increasing after 1995. In contrast, the Indochina Peninsula (Region 3) witnessed an evident turning point in 1995, with the impact of altitude on precipitation increasing initially and then decreasing. For the Indian Peninsula (Region 4), the volatility of average $\omega_1$ significantly increased after 1995, with higher overall values than before 1995. Southeast China (Region 2) and inland regions (Region 6) are relatively unaffected by the turning point in 1995, with the average $\omega_1$ showing significant downwards and upwards trends, respectively. Analysis of the other coefficients further reinforces that the terrain-precipitation relationship is intricately linked to climate change (Supplementary Information S1.5).

This analysis shows that the terrain-precipitation relationship is influenced by climate change, and different regions exhibit significantly different change patterns. The observed "1995 turning point" aligns with other climate change phenomena since the 1990s, such as the acceleration of warming in China(*27*), changes in the Arctic Oscillation (AO) and North Atlantic Oscillation (NAO) indexes in high latitude regions(*28*, *29*), and the weakening of the wind-driven south equatorial current(*30*). These findings offer a novel perspective on global climate change and suggest that climate change may have far-reaching implications for the terrain-precipitation relationship, which has been overlooked previously.

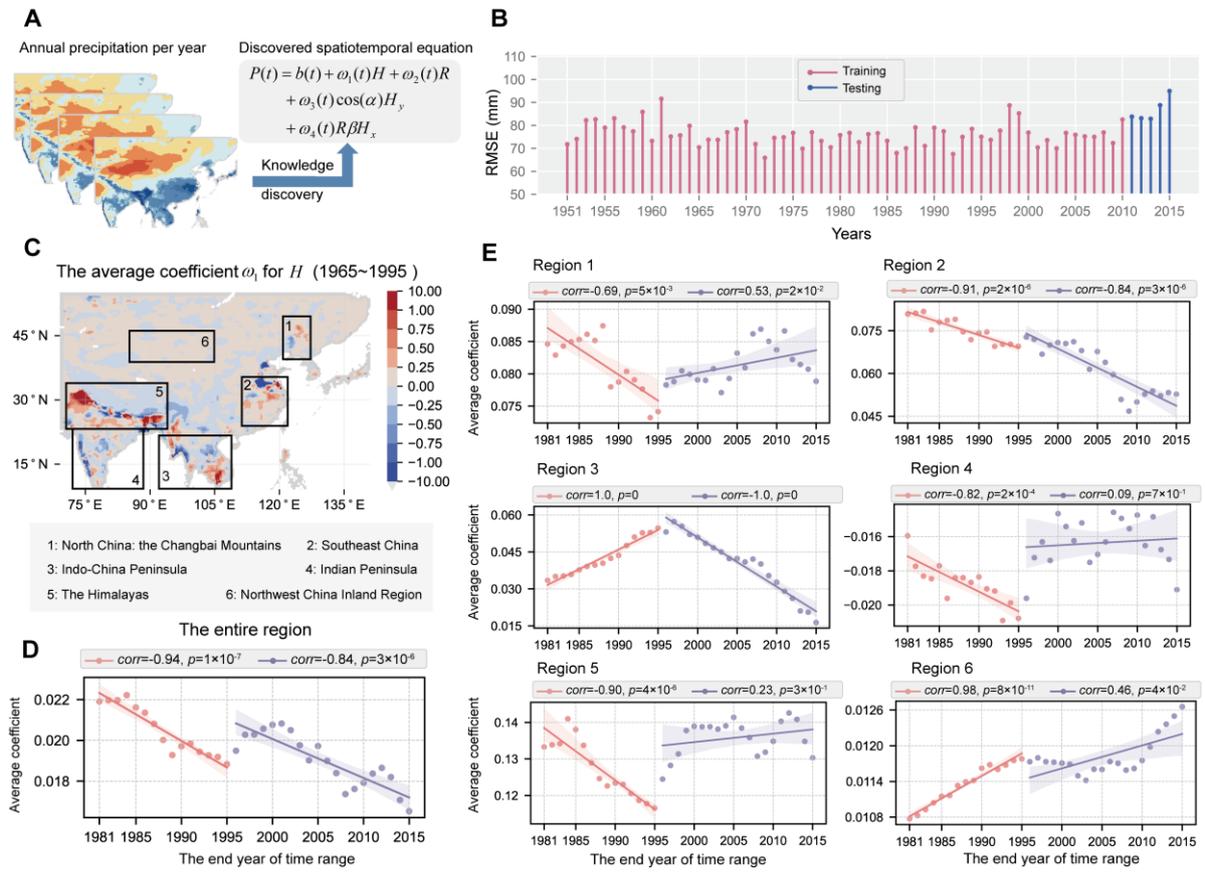

**Fig. 3. The discovered spatial-temporal formula and the influence of climate change on the terrain-precipitation relationship.** **(A)** Sketch map for discovering the spatiotemporal formula from annual precipitation data. **(B)** The root mean squared error (RMSE) of the discovered spatiotemporal formula to fit the annual precipitation in each year. **(C)** Several characteristic regions were selected to provide a comprehensive illustration of the impact of climate change on the terrain-precipitation relationship across diverse regions, including the Changbai Mountains in North China, Southeast China, the Indochina Peninsula, the Indian Peninsula, the Himalayas, and the inland region of Northwest China. Information on these characteristic regions is provided in Supplementary Information S1.5. **(D)** The trend of the average normalized coefficient $\omega_1$ in the entire region. **(E)** The trend of the average normalized coefficient $\omega_1$ in the investigated characteristic regions. The time range is 30 years.

**Decoupling the impacts of terrain and climate variables on precipitation**

Considering the intricate relationship between climate variables and terrain-precipitation relationships, simple regression models are not sufficient to capture the complexities. Hence, we employed a physics-informed neural network (PINN, details refer to *Methods*) to model this relationship. The network takes the 23 climate variables as input and predicts the coefficients $\omega$ and $b$ in Eq. (2) as output. These variables are obtained from correlation analysis and clustering (Fig. 4A) of 33 relevant variables (Table S2) from ERA5. To enhance the model's accuracy, we integrated Eq. (2) as a physical constraint into the loss function during training. During the training phase, we utilized data from 1951 to 2010 as the training set and data from 2011 to 2015 as the testing set. After successfully training the model, we used the climate variables from the testing set to predict the coefficients $\omega$ and $b$, thereby utilizing Eq. (2) to directly predict precipitation patterns from 2011 to 2015, which can be represented as follows:

$$\hat{P} = \hat{\omega}_1(X_c)H + \hat{\omega}_2(X_c)R + \hat{\omega}_3(X_c)\cos(\alpha)H_y + \hat{\omega}_4(X_c)(t)R\beta H_x + \hat{b}(X_c), \quad (3)$$

where $\hat{P}$ denotes the predicted precipitation, $X_c$ is the matrix of 23 climate variables, and $\hat{\omega}$ and $\hat{b}$ are the obtained coefficients from the output of the trained PINN with $X_c$ as the input. Importantly, the previous GWR regression fitting based on Eq. (2) could only explain known years instead of predicting future trends since it requires observed precipitation to obtain the coefficient field. In contrast, our developed prediction model, Eq. (3), can directly forecast the patterns of future annual precipitation based on future climate and known terrain variables since the coefficient field is directly predicted from climate variables. By converting Eq. (2) into Eq. (3), we successfully decouple the effects of climate and terrain variables on precipitation, thereby transforming a previously time-varying model into a climate-varying model. As depicted in Fig. 4B, our prediction model exhibits satisfactory accuracy in forecasting changes in precipitation patterns.

Unlike the AI large models used in previous literature(*8, 21*), our prediction model offers a distinctive advantage by being expressible in an explicit formula. Each term in this formula possesses a clear physical meaning, granting the model superior interpretability, which enables deeper insights into the underlying mechanisms governing precipitation patterns in response to climate and terrain factors. In Eq. (3), the coefficients of the terrain term are predicted by climate variables using PINN. Although PINN itself is a black-box model, through an interpretable machine learning technique called SHapley Additive exPlanations (SHAP)(*31*), we can explain the influence of each climate variable on the terrain-precipitation relationship

(for details, refer to *Methods*). The Shapley values of the coefficients are illustrated in Fig. 4C, which shows six climate variables with the most pronounced impact on each coefficient.

Notably, we observed that climate variables related to vegetation, such as vegetation cover and leaf area index, play a crucial role in shaping the terrain-precipitation relationship. Interestingly, the influence of high and low vegetation on precipitation patterns diverges. The high vegetation cover (*cvh*) mainly affects the constant *b*, which characterizes precipitation brought by climate zones, while low vegetation cover (*cvl*) affects both the constant term and terrain-related terms. On the one hand, more high vegetation and less low vegetation will enhance the influence of the constant *b* on precipitation. On the other hand, more low vegetation will enhance the impact of terrain-related terms, such as altitude and relief on precipitation.

Additionally, traditional meteorological indicators, such as temperature, wind speed, surface pressure, and cloud cover, are discovered to influence the terrain-precipitation relationship. However, their impact on individual terrain parameters varies. For instance, temperature significantly impacts all terrain-related variables, cloud cover has a pronounced effect on the altitude-precipitation relationship, and surface pressure exerts a notable influence on relief and altitude gradient terms. By elucidating the specific climate variables and their distinct impacts on the terrain-precipitation relationship, our findings deepen the understanding of the intricate mechanisms governing regional precipitation patterns.

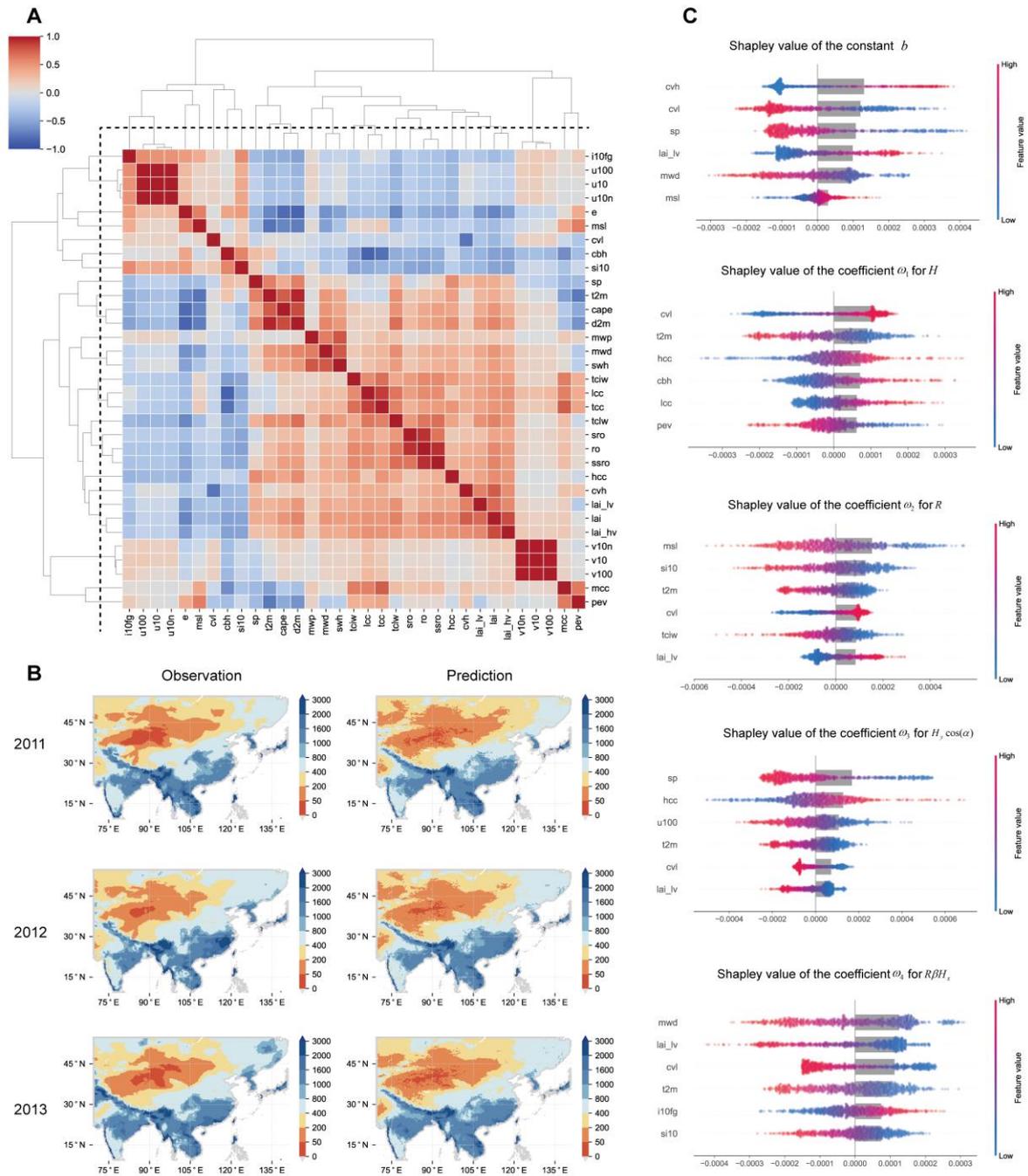

**Fig. 4. Analysis of the influence of climate variables on the terrain-precipitation relationship. (A)** The clustering correlation matrix for the 33 related climate variables. The dotted line is the selection criterion. **(B)** The observed and predicted annual precipitation fields from our prediction model. **(C)** The Shapley values of climate variables for each coefficient. The grey bar is the absolute mean of the Shapley values, and a longer bar indicates greater importance. Here, the 6 variables with the largest importance are displayed.

**Long-term prediction of high-resolution future precipitation patterns**

The high-resolution long-term prediction of precipitation models remains a challenge. As of now, the Coupled Model Intercomparison Project Phase 6 (CMIP6)(*32*) model data have been instrumental in simulating and evaluating changes in precipitation patterns until 2100 across various climate scenarios. While CMIP6 provides valuable insights into large-scale climate projections, its long-term predictions are limited by their low resolution (1°× 1°), which restricts its ability to find intricate changes at a detailed level. Fortunately, the proposed prediction model provides a path to obtain a long-term prediction of high-resolution precipitation from low-resolution future climate variables in CMIP6. According to Eq. (3), the prediction requires both terrain and climate data. Benefiting from the high-resolution terrain terms that are spatial fields that remain constant over time, even with low-resolution climate variables, a fine-scale precipitation prediction can still be obtained from Eq. (3). Despite potential accuracy trade-offs with low-resolution climate data, our model remains reliable and interpretable, making it a valuable tool for predicting precipitation patterns under varying climate scenarios.

In this section, we take the 2-metre air temperature (*t2m*) and leaf area index (*lai*) as examples to demonstrate the long-term precipitation prediction ability of our prediction model under different climate scenarios, including SSP126, SSP370 and SSP585, which utilize the shared socioeconomic pathway (SSP) to characterize the impact of human activities in future scenarios(*33*). Specifically, SSP585 represents a future scenario characterized by unsustainable practices and careless human behaviour, leading to catastrophic consequences. In contrast, SSP126 portrays a future scenario driven by responsible and restrained human actions, aiming to mitigate the impacts of climate change. Unlike CMIP6, which directly predicts future precipitation under various climate scenarios, our experiment took a different approach. We specifically focused on exploring the isolated impact of individual climate variables by keeping other climate parameters constant at their values in 2015 while altering the specified variable. This allowed us to investigate how changes in a single climate factor under different climate scenarios influence future precipitation patterns.

Here, the difference between the average annual precipitation predicted for the period from 2071 to 2100 and the period from 1986 to 2015 is calculated. The low-resolution *t2m* values provided by CMIP6 under different climate scenarios are depicted in Fig. 5A. The predicted high-resolution variation field of annual precipitation caused by the change in *t2m* under the corresponding climate scenarios is illustrated in Fig. 5B. Our findings reveal that under the high forcing climate scenario (i.e., SSP 370 and SSP585) where *t2m* is higher, precipitation presents an evident increase in southeastern China and the Indian peninsula. However, in some regions with complex terrain characteristics, it leads to reduced precipitation, notably observed in the northern foothills of the Himalayas and Northern Indo-China Peninsula. This finding highlights how the impact of the increase in *t2m* on precipitation is influenced by the terrain. Similarly, we also examined the impact of changes in *lai* under different climate backgrounds on precipitation. As previously revealed, vegetation significantly influences the terrain-precipitation relationship. Fig. 5C demonstrates that under the high-forcing climate background, the leaf area index experiences an increase. Unlike the impact of temperature, the effect of vegetation on precipitation is mainly concentrated over the Tibetan Plateau and West China.

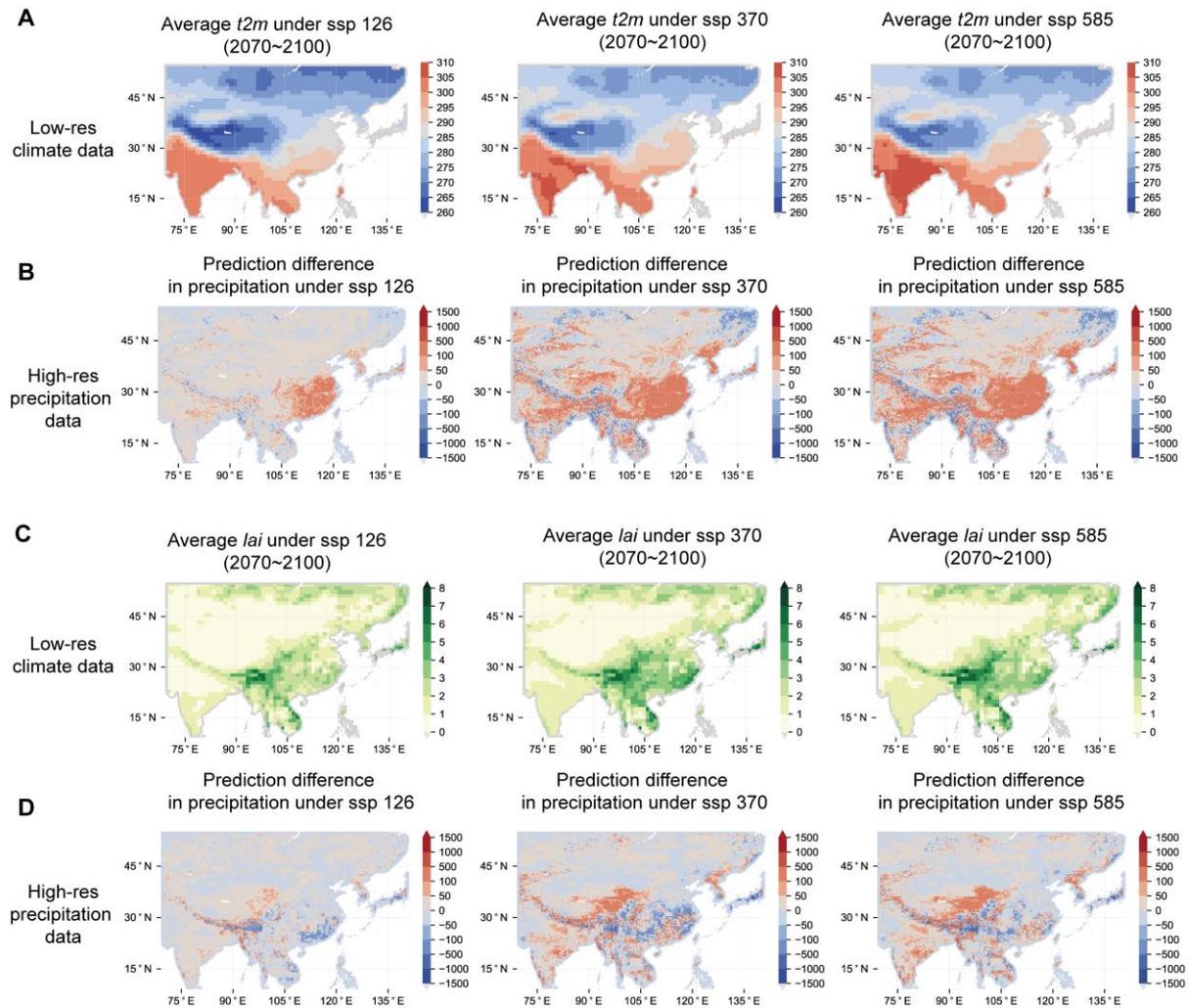

**Fig. 5. Long-term precipitation prediction under different climate scenarios. (A)** The low-resolution 2-metre air temperature (*t2m*) variable values are provided by CMIP6 under different climate scenarios. **(B)** The predicted high-resolution variation field on average annual precipitation is caused by the change in *t2m* under the corresponding climate scenarios. **(C)** The low-resolution leaf area index (*LAI*) values provided by CMIP6 under different climate scenarios. **(D)** The predicted high-resolution variation field on average annual precipitation caused by the change in *lai* under the corresponding climate scenarios.

**Discussion**

While deep learning models have made notable progress in precipitation forecasting(*6–8*), these black-box models lack interpretability, limiting their capacity to promote the understanding of underlying mechanisms. In contrast, explicit equations provide a formalized framework for capturing the fundamental principles underlying the phenomenon. Historically, equations were manually derived under simplified assumptions, which becomes impractical when dealing with intricate terrain-precipitation dynamics. Geographical regression models provide an alternative method, but they still rely on physical explanatory terms proposed by human researchers(*18–20*). In this context, the proposed AI-driven knowledge discovery framework becomes a powerful tool to revolutionize terrain-precipitation understanding by automatically unearthing explicit equations that were previously undiscovered.

While conventional knowledge discovery methods have focused on uncovering partial differential equations (PDEs), they have primarily remained in the proof-of-concept stage, rediscovering known governing equations to confirm the effectiveness of methods(*23*). In contrast, our proposed GA-GWR approach advances beyond the proof-of-concept stage by amalgamating geoscience techniques to explore new knowledge in practice. The discovered terrain-precipitation formula can be interpreted physically, which not only enhances the understanding of precipitation but also enables high-resolution forecasting.

In summary, our study used AI-driven knowledge discovery techniques to uncover explicit formulas governing the terrain-precipitation relationship. Our findings elucidate that this relationship is dynamic, undergoing changes with evolving climate patterns, as exemplified by the '1995 turning point' phenomenon. This relationship in regions characterized by intricate terrain complexity presents divergent trends before and after the mid-1990s. This intricate interplay suggests that the impact of climate change on the terrain-precipitation relationship may be more profound than previously anticipated, which has a far-reaching influence on the future precipitation distribution. Through this formula, accurate and fine-scale precipitation predictions are conducted, even with low-resolution future climate data. Such predictions enable a systematic analysis of climate change impacts on annual precipitation across diverse terrains, providing invaluable insights into evolving precipitation patterns under future climate scenarios. Our study predominantly focused on precipitation patterns at the mesoterrain scale, but local microclimatic influences within small-scale terrains disappeared in the grid point averaging process, which deserves future research to unearth explicit equations and elucidate the underlying mechanisms behind these phenomena.

**Materials and Methods**

**Dataset**

For the AI-empowered discovery of the terrain-precipitation relationship, the Asian terrain and precipitation data are obtained from the Asian Precipitation-Highly Resolved Observational Data Integration Toward Evaluation of Water Resources Project (APHRODITE) dataset. This dataset provides a high-quality gridded precipitation dataset for East Asia, and the dataset was obtained from rain gauge observation records covering more than 50 years with high spatial and temporal resolutions (*24*), and it has been applied in studies of precipitation over contiguous China (*34, 35*). For the precipitation data, we use its daily precipitation product APHRO_MA at a $0.25° \times 0.25°$ spatial resolution from 1951 to 2015, which contains 184 grids in the latitude direction and 288 grids in the longitude direction. The daily precipitation is accumulated each year to obtain annual precipitation. For the terrain data, four variables in the APHRODITE dataset are considered, including altitude, relief, slope, and aspect.

For the climate-terrain-precipitation formula discovery, the climate variables from the fifth generation of the European Centre for Medium-Range Weather Forecasts (ECMWF) reanalysis, abbreviated as ERA5, are employed, combining model data with observations from across the world into a globally complete and consistent dataset using the laws of physics(*36*). Specifically, the ERA5 monthly averaged data on single levels at a $0.25° \times 0.25°$ spatial resolution from 1951 to 2015 are utilized. In the ERA5 dataset, there are hundreds of climate variables that estimate atmospheric, ocean-wave and land-surface quantities. In this work, 33 variables are selected from the ERA5 dataset, which is regarded as relevant to precipitation based on experience and domain knowledge. The names and introductions of these variables are detailed in Table S2. To obtain the annual data, these variables are accumulated for each year. According to the clustering correlation matrix in Fig. 3, similar variables are dropped in the feature engineering, and 23 variables are ultimately selected to learn the influence of climate on the terrain-precipitation relationship (Table S2).

In the long-term prediction of high-resolution future precipitation patterns under different climate scenarios, the Coupled Model Intercomparison Project Phase 6 (CMIP6)(*32*) model data are utilized, which brings together a diverse community of climate researchers, modelling centres, and institutions from around the world to simulate and analyse the complex interactions of Earth's atmosphere, oceans, land, and cryosphere. Benefiting from CMIP6, researchers can simulate and predict future climate conditions under various scenarios of greenhouse gas emissions and other external forcings (*37, 38*). In this work, the monthly simulated data for 2 m temperature (*t2m*) and leaf area index (*lai*) at a $1°\times 1°$ spatial resolution from 1840 to 2100 are downloaded. The monthly data are averaged for each year to obtain the annual data. Three climate scenarios are considered, including SSP126, SSP370 and SSP585.

**Artificial intelligence knowledge discovery**

**Backgrounds and problem settings.** Historically, the discovery of new knowledge has been a slow process, demanding years of arduous exploration and analysis. Galileo spent 17 years fitting models and making calculations using Tycho Brahe's more than three-decade observations of Mars' orbit to establish three laws of planetary motion. Subsequently, Newton extended these findings theoretically, culminating in the formulation of his law of universal gravitation. This lengthy process, involving three brilliant scientists over nearly a century, underscores the time-intensive nature of knowledge discovery.

In the modern era, innovative data analysis techniques, such as artificial intelligence (AI) and machine learning, have revolutionized knowledge discovery. These tools can quickly analyse extensive datasets, uncover intricate patterns, and reveal hidden relationships, significantly expediting the pace of advancements across various fields. Knowledge discovery is a potent process that combines data mining, machine learning, and domain expertise. It involves various methods to extract valuable insights and uncover hidden relationships in large datasets. Its primary aim is to reveal new knowledge and patterns, enhancing our understanding of phenomena using AI techniques.

In this work, we proposed an AI-driven knowledge discovery framework called Genetic Algorithm-Geographic Weighted Regression (GA-GWR) to discover the underlying terrain-precipitation relationship based on the field knowledge of geoscience. The GA-GWR approach combines two essential components: the genetic algorithm (GA) as an advanced optimization technique and geographic weighted regression (GWR) as a powerful descriptive method in geoscience. Conventional GWR methods typically necessitate predetermining the relevant terms in the regression model, which relies heavily on experience. By leveraging the capabilities of the GA, GA-GWR efficiently explores a large search space of possible formulas, enabling it to identify the most appropriate combination of terms for the study area. This dynamic feature selection process significantly enhances the accuracy and efficiency of the model. The problem settings can be described as follows:

$$P = \Phi(H, R, \alpha, \beta, \text{K}, H^2 R, R\beta \cos(\alpha), ...) \cdot \vec{\xi}, \tag{4}$$

where $\Phi$ is an operator that indicates a linear combination of terms; $P$ is the observed precipitation data; and $\vec{\xi}$ is the coefficient field. The content of the operator $\Phi$ is numerous potential terms combined with terrain variables. For knowledge discovery, it aims to identify nonzero terms and corresponding coefficients from data. Therefore, the difficulty of the problem lies in how to consider both the accuracy and parsimony of the discovered formula. The details of the process of GA-GWR are described below.

**Geographic weighted regression.** Geographic weighted regression (GWR) is an advanced spatial analysis technique. It differs from traditional global regression models by acknowledging that relationships between variables can vary across space due to local factors and spatial heterogeneity. GWR involves separate regressions at different locations within the study area, capturing the unique spatial patterns and variations in the data. This adaptation is completed by giving different weights to nearby data points and considering their impact on the regression at each specific location. The GWR model can be written as follows:

$$P_i = b(u_i, v_i) + \sum_{k=1}^{m} \omega_k(u_i, v_i) x_{i,k} + \varepsilon_i, i = 1, 2, ..., n, \tag{5}$$

where $i$ is the grid index; $u_i$ and $v_i$ are the longitude and latitude of grid $i$, respectively; $x_{i,k}$ is the value of the $k^{th}$ term at grid $i$; $b$ and $\omega_k$ are the spatially varying fields of coefficients; $\varepsilon_i$ is the error; $P_i$ is the observed precipitation data; $m$ is the number of independent variables; and $n$ is the total number of grids.

The parameter estimation of geographic regression models is similar to that of the general linear regression model. However, due to the unique spatial characteristics of geographic data, the presence of spatial dependence and heterogeneity in the data leads to many variables, reaching $n \times (m+1)$, leading to a nonparametric regression. To ensure accurate parameter estimation, the foundational principle in geography known as Tobler's first law is adopted, which asserts that "everything is related to everything else, but near things are more related than distant things." Guided by this spatial principle, the weighted least squares method is adopted for estimating the regression coefficients. In the conventional GWR model, the weight function is a function that decreases with the distance $d_{ij}$ between centre point $i$ and another point $j$, which is written as follows:

$$w(u_j, v_j) = f_\theta(d_{ij}), \qquad (6)$$

where $w(u_j, v_j)$ is its weight of point $j$; $f_\theta$ is the weight function, which usually employs the Gaussian function. However, Eq. (6) requires calculating the distance between each other point and the centre point, which results in a significant computational burden and low efficiency of the GWR model. Therefore, in practical use, local linear GWR regression with stronger operability is commonly used for coefficient estimation (Fig. S5D). Considering that the involved dataset is grid data, for each central grid $(u_i, v_i)$, the influence domain is a square centred around on $(u_i, v_i)$ with $2 \times wl$ as the side length, where $wl$ refers to the maximum window length. Therefore, the weight for the estimation of the coefficients in this work can be represented as follows:

$$w(u_j, v_j) = \begin{cases} 1, & wl_j \leq wl \\ 0, & wl_j > wl \end{cases}, \qquad (7)$$

where $wl_j$ refers to the window length of point $j$ to centre point $i$. This finding means that only grids in the influence domain are involved in the parameter estimation of the centre point. Therefore, the estimation of the local coefficients can be written as follows:

$$[\hat{b}_i, \hat{\omega}_i] = (X_i^T X_i)^{-1} X_i^T \vec{P}_i, \qquad (8)$$

$$X_i = \begin{bmatrix} 1 & x_1^{u_{i-wl}, v_{i-wl}} & x_2^{u_{i-wl}, v_{i-wl}} & \cdots & x_k^{u_{i-wl}, v_{i-wl}} \\ 1 & x_1^{u_{i-wl+1}, v_{i-wl}} & x_2^{u_{i-wl+1}, v_{i-wl}} & \cdots & x_k^{u_{i-wl+1}, v_{i-wl}} \\ \cdots & \cdots & \cdots & \cdots & \cdots \\ 1 & x_1^{u_{i+wl}, v_{i+wl}} & x_2^{u_{i+wl}, v_{i+wl}} & \cdots & x_k^{u_{i+wl}, v_{i+wl}} \end{bmatrix}, \qquad (9)$$

$$\vec{P_i} = \begin{bmatrix} P^{u_{i-wl},v_{i-wl}} \\ P^{u_{i-wl+1},v_{i-wl+1}} \\ ... \\ P^{u_{i+wl},v_{i+wl}} \end{bmatrix}, \tag{10}$$

where $X_i$ is the matrix of independent variables (terrain terms), $\vec{P_i}$ is the vector of the dependent variable (precipitation data), and $\vec{b_i}$ and $\vec{\omega_i}$ are the estimated coefficients for grid $i$. Therefore, the fitted precipitation $\hat{P}_i$ at grid $i$ can be calculated as follows:

$$\hat{P}_i = \vec{b}_i(u_i, v_i) + \sum_{k=1}^{p} \vec{\omega}_k(u_i, v_i) x_{i,k}, i = 1, 2, ..., n. \tag{10}$$

**GA-GWR.** The genetic algorithm is a powerful optimization technique inspired by natural selection and evolution. It consists of several essential steps, including digitization, crossover, mutation, fitness calculation, and evolution. To address the specific challenges of the concerned research problem in this work, we have made crucial modifications to the conventional genetic algorithm, such as adding the mechanism of distinction, to handle the complexities encountered in this work.

In this study, we introduce a versatile digitization technique that allows for clear and flexible expression of potential terms. The principle of our digitization method is depicted in Fig. S5A. The formula is digitized into genomes composed of gene modules. The number in the gene modules refers to the index of basic terrain variables, which are called basic genes. For example, 0 refers to the altitude $H$, and 2 refers to the relief $R$. The term (or gene module) is constructed with the basic genes. For example, [0,2] refers to $H \times R$. Finally, several terms consist of the formula by addition. For instance, {[0], [0,3], [0,1,2]} refers to $H + H\alpha + H\beta R$. With this digitization principle, each potential relationship corresponds to a specific genome, which paves the way for the subsequent process. For the initial generation, a few genomes are randomly generated.

The genetic algorithm involves two key steps: crossover and mutation. During crossover, gene modules are exchanged between parent genomes to create children, exploring new possibilities (Fig. S5C). Afterwards, mutation further diversifies the population using three methods: delete-module, add-module, and basic-gene mutations (Fig. S5B). The delete module removes a random module, the add module adds a new random module, and the basic gene mutation replaces a basic gene with a new random gene. These mutations enhance the algorithm's ability to explore different combinations and widen the search area. By using crossover and mutation, GA-GWR efficiently navigates the solution space, increasing the chances of discovering suitable mathematical representations.

In the fitness calculation step, each genome's fitness is calculated to measure the quality of the genome, which is defined as follows:

$$F = MSE + l_0\_penalty \cdot m, \tag{11}$$

with

$$MSE = \frac{1}{n}\sum_{i=1}^{n}(P_i - \hat{P}_i)^2, \tag{12}$$

where $F$ denotes the fitness, which consists of the $MSE$ part and $l_0$ penalty term; $MSE$ is calculated according to Eq. (12), where $m$ is the number of terms in the formula, and $n$ is the number of grids. The $\hat{P}_i$ and $P_i$ are the fitted and observed precipitations from GWR. Here, a smaller fitness indicates a better genome (i.e., formula).

Once the fitness has been computed for all genomes, a selection process is employed to determine which genomes will form the next generation. Half of the genomes with the smallest fitness values are retained, while the other half are replaced with newly generated random genomes. This selection and replacement mechanism ensures that promising solutions persist while introducing fresh gene modules to maintain diversity. In cases where the optimization process reaches a local optimum and the optimal structure remains unchanged across consecutive generations, we employ an extinction mechanism to assist in breaking free from local optima. During the extinction process, only the first three optimal structures are maintained unchanged, while all other genomes are dropped. To simulate the concept of extinction in nature, new genomes are generated based on the first three optimal structures to replace the dropped genomes. This approach mimics the survival and reproduction of the most adaptable species considering environmental challenges. By promoting the survival of the most promising solutions and discarding less favourable ones, the extinction mechanism encourages a more refined optimization process even when trapped in local optima.

The evolutionary process continues iteratively until the maximum specified number of iterations is reached. At the end of the evolution, the optimal structure is determined by identifying the genome with the smallest fitness value in the final generation. This genome represents the best-discovered formula, providing the most suitable mathematical representation for target knowledge discovery.

**Discovery of terrain-precipitation formula via GA-GWR**

In this work, the GA-GWR framework is utilized to identify the terrain-precipitation formula. The dependent variable is the average annual precipitation, while the independent variables are terrain-related variables, including altitude ($H$), relief ($R$), slope ($\alpha$), and aspect ($\beta$). The variables are standardized to facilitate the process of GA-GWR. Eleven basic genes are defined to generate genomes, including $H$ (0), $\beta$ (1), $R$ (2), $\alpha$ (3), $H_x$ (4), $H_y$ (5), $\sqrt{H}$ (6), $\sin(\alpha)$ (7), $\cos(\alpha)$ (8), $\sin(\beta)$ (9), and $\cos(\beta)$ (10). Notably, the basic genes $H$ and $\sqrt{H}$ are employed to generate terms in decimal order (e.g., $H^{1.5}$). The population and maximum generation of GA-GWR are 200 and 200, respectively, and $l_0\_prenalty$ is $10^{-5}$. The $wl$ for calculating the fitness by GWR is 3. The influence of these parameters is discussed in Supplementary Information S1.3 Considering that there are many default values located on the ocean in the dataset, only central points with more than 5 valid values in the influence domain are considered. Therefore, there are 32,740 available central points in the dataset (approximately 61.8% of the total grids). The GA-GWR is conducted three times with different random seeds to generate the initial genomes to avoid the local minimum, and the evolutionary process that generates the best genome with the smallest fitness is provided in Table S3.

**Discovery of the spatial-temporal terrain-precipitation formula**

The discovery of the spatial-temporal terrain-precipitation formula, Eq. (2), is similar to the above procedure, but the calculation of fitness is different. Here, the annual precipitation from 1951 to 2010 (65 years) is the observed data. Our knowledge discovery aims to discover a spatial-temporal formula that is most suitable to describe the annual precipitation in each year.

Therefore, the GWR model is revised as follows:

$$F = \frac{1}{n_t}\frac{1}{n}\sum_{q=1}^{n_t}(P_i^q - \hat{P}_i^q)^2 + l_0\_penalty \cdot m, \qquad (13)$$

where $n_t$ is the number of years; $P_i^q$ and $\hat{P}_i^q$ are the observed and fitted precipitation at grid $i$ in year $q$. The $\hat{P}_i^q$ is obtained from the GWR in the same manner as Eq. (8), with the dependent variable being the annual precipitation in year $q$. The GA-GWR is conducted three times with different random seeds to generate the initial genomes to avoid the local minimum, and the evolutionary process that generates the best genome with the smallest fitness is provided in Table S4. To test the performance of the discovered spatial-temporal terrain-precipitation formula, the annual precipitation from 2011 to 2015 (5 years) is taken as testing data.

**Decoupling the influence of climate and terrain on precipitation**

The previously discovered formulas, Eq. (1) and Eq. (2), need GWR to determine the coefficient fields, which means that the function of these formulas is constrained to explain the precipitation and lead to the discovery of the terrain-precipitation relationship. In this work, a physics-informed neural network (PINN) is employed to decouple the influence of climate and terrain on precipitation. In Eq. (2), it is obvious that the influence of terrain is described by terrain-related terms, while the coefficients vary spatiotemporally and are mainly governed by climate factors. In the PINN, the inputs are 23 relevant climate variables obtained from the ERA5 dataset, and the outputs are 5-dimensional vectors that contain the coefficients $\omega_i$ and $b$. To enhance the model's accuracy, we integrated Eq. (2) as a physical constraint into the loss function during training. The loss function is written as follows:

$$Loss = \lambda_{Data}Loss_{Data} + \lambda_{Formula}Loss_{Formula}, \qquad (14)$$

$$Loss_{Data} = \frac{1}{N_{train}}\frac{1}{m}\sum_{k=1}^{m}\sum_{j=1}^{N_{train}}(y_j^k - \hat{y}_j^k)^2, \qquad (15)$$

$$Loss_{Formula} = \frac{1}{N_{train}}\sum_{j=1}^{N_{train}}(P_i - \hat{P}_i)^2. \qquad (16)$$

Here, $Loss_{Data}$ is the loss between the observed coefficient $y_j^k$ and predicted coefficients $\hat{y}_j^k$, where $N_{train}$ is the number of training data and $m$ is the number of terms. $Loss_{Formula}$ is the loss between the observed precipitation $P_i$ and predicted precipitation $\hat{P}_i$ according to Eq. (3) utilizing the predicted coefficients. In this work, $\lambda_{Data} = 1$ and $\lambda_{Formula} = 10$. During the training phase, we utilized data from 1951 to 2010 as the dataset, which is split into 80%, 10%, and 10% as the training set, validating set, and testing set, respectively. The data from 2011 to 2015 are taken as the out-of-sample testing set. For the PINN, there is one input layer with 23 neurons, one output layer with 5 neurons, and 5 hidden layers with 256 neurons. The activation function is LeakyReLU. The number of training epochs is 4,000.

**SHapley Additive exPlanations (SHAP)**

The principle of SHapley Additive exPlanations (SHAP) is to provide a unified framework for explaining the output of any machine learning model by allocating the contribution of each feature to the final prediction(*31*). It is based on cooperative game theory and ensures that the explanations satisfy certain desirable properties, such as fairness, consistency, and local accuracy (*39*). The Shapley values are obtained from SHAP to measure the contributions of variables, which consider all possible combinations of features and their contributions to the final prediction. It ensures that the explanation is both fair and consistent across different feature combinations. The calculation formula of the Shapley value of component *i* in all collections of participants *I* can be written as follows:

$$\varphi_i = \sum_{s \in S_i} \frac{(|s|-1)!(n-|s|)!}{n!} \delta_i(s), \tag{17}$$

$$\delta_i(s) = v(s) - v(s \setminus \{i\}), s \in S_i. \tag{18}$$

Here, $\varphi_i$ is the SHAP value of participant *i*; $S_i$ is the set of all possible combinations including member *i*; $s\setminus\{i\}$ is the new set produced by removing member *i* from *s*; *v* is the contribution; and $w(|s|)$ is the weight. *N* is the number of features in the input data. Therefore, the process of calculating the Shapley value can be roughly summarized as calculating the average value of marginal contribution in nodes of *n*! situations in all sequence permutations. In practice, calculating Shapley values for all features can be computationally expensive since it involves evaluating the model for all $2^n$ possible coalitions of features. Various approximation methods and techniques have been developed to efficiently compute Shapley values for complex models.

**Data availability**

The dataset generated in this study has been deposited in the Github repository, https://github.com/woshixuhao/Discovery_of_terrain_precipitation_formula/tree/main.

**Code availability**

All original code has been deposited in the Github repository, https://github.com/woshixuhao/Discovery_of_terrain_precipitation_formula/tree/main.

**Acknowledgements**

This work was supported and partially funded by the National Natural Science Foundation of China (Grant No. 52288101), the National Center for Applied Mathematics Shenzhen (NCAMS), the Shenzhen Key Laboratory of Natural Gas Hydrates (Grant No. ZDSYS20200421111201738), and the SUSTech – Qingdao New Energy Technology Research Institute.

**Author contributions**

Y.C. and D.Z. supervised the whole project. L.L., and J.L. provided the precipitation dataset. H.X. analysed the data, proposed the algorithm, conducted numerical experiments, and wrote the manuscript. H.X. and Z.Z. analysed the results and obtained the conclusion. Y.C. conceived the idea and designed the overall research.

**Competing interests**

All authors declare that they have no competing interests.

# Supplementary Information for

# Interpretable AI-Driven Discovery of Terrain-Precipitation Relationships for Enhanced Climate Insights

Hao Xu[1], Yuntian Chen[2,*], Zhenzhong Zeng[3], Nina Li[4], Jian Li[5,*], and Dongxiao Zhang[2,6,*]

[1] BIC-ESAT, ERE, and SKLTCS, College of Engineering, Peking University; Beijing 100871, P. R. China

[2] Ningbo Institute of Digital Twin, Eastern Institute of Technology; Ningbo, Zhejiang 315200, P. R. China

[3] South University of Science and Technology of China; Shenzhen, Guangdong 518055, P. R. China

[4] National Meteorological Center, China Meteorological Administration; Beijing 100081, China

[5] Chinese Academy of Meteorological Sciences; Beijing 10081, China

[6] Department of Mathematics and Theories, Peng Cheng Laboratory; Shenzhen 518000, Guangdong, P. R. China

# 1. Supplementary text

## 1.1 Preprocessing the dataset

In this work, the precipitation data and four terrain variables, including altitude, relief, slope, and aspect, are obtained from the Asian Precipitation-Highly Resolved Observational Data Integration Toward Evaluation of Water Resources Project (APHRODITE) dataset, which are depicted in Fig. S1. Due to the substantial differences in the order of these variables, it becomes imperative to standardize them prior to knowledge discovery by genetic algorithm-geographic weighted regression (GA-GWR). The max-min standardization method is adopted here.

$$P = \frac{P - P_{min}}{P_{max} - P_{min}}, \tag{S.1}$$

$$R = \frac{R - R_{min}}{R_{max} - R_{min}} = \frac{R}{R_{max}}, \tag{S.2}$$

$$H = \frac{H - H_{min}}{H_{max} - H_{min}}. \tag{S.3}$$

$P_{min}$, $R_{min}$, and $H_{min}$ are the minimum precipitation, relief, and altitude, respectively, and $P_{max}$, $R_{max}$, and $H_{max}$ are the maximum precipitation, relief, and altitude, respectively. Since $R_{min}=0$, Eq. (S.2) is simplified. Moreover, as the slope and aspect are described in the form of degrees, we convert them to radians, which is written as follows:

$$\alpha = \frac{\pi}{180}\alpha, \tag{S.4}$$

$$\beta = \begin{cases} \dfrac{\pi}{180}\beta, & \beta \leq 180° \\ \dfrac{\pi}{180}(360 - \beta), & 180° < \beta \leq 360° \end{cases}. \tag{S.5}$$

Here, aspect $\beta$ is preprocessed to be the radian angle to the north direction.

In this work, the gradients of the altitude at the grid in the direction of longitude and latitude are also considered to parameterize the terrain, which is calculated as follows:

$$H_x(u_i, v_j) = \frac{H(u_{i+1}, v_j) - H(u_{i-1}, v_j)}{(u_{i+1} - u_{i-1})\Delta d \cos(v_j)}, \tag{S.6}$$

$$H_y(u_i, v_j) = \frac{H(u_i, v_{j+1}) - H(u_i, v_{j-1})}{(v_{j+1} - v_{j-1})\Delta d}, \tag{S.7}$$

where $u$ and $v$ are the longitude and latitude, respectively; $i$ and $j$ are the index of the grid; $\Delta d$ is the distance of 1° on the equator (111 km in this work); and $H_x$ and $H_y$ are the gradient in the direction of longitude and latitude, respectively.

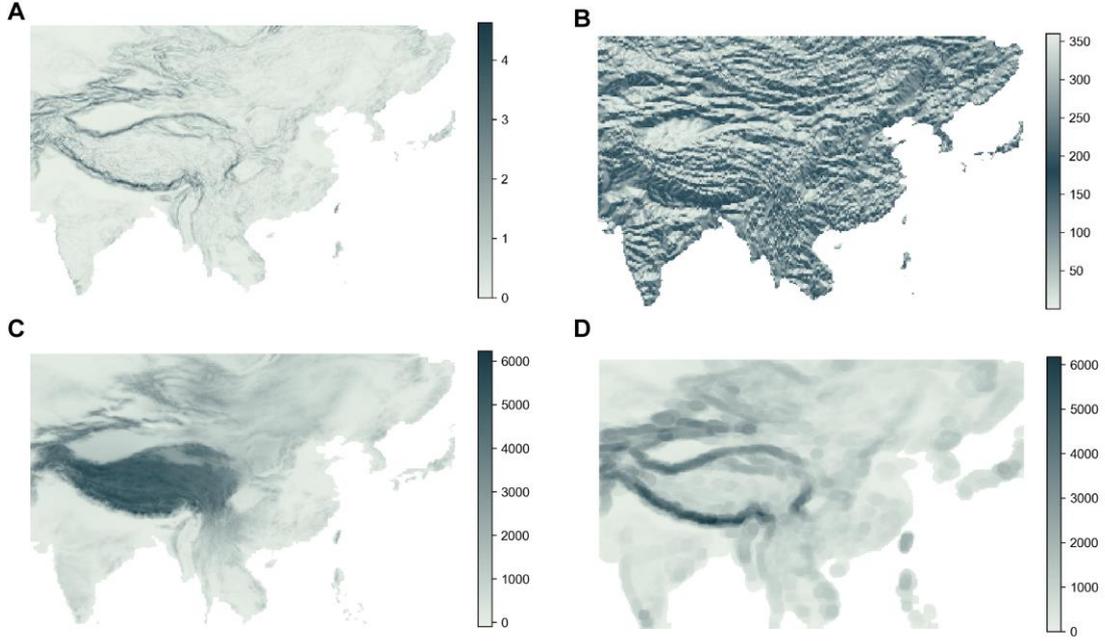

**Fig. S1. The visualization of four terrain variables in the dataset. (A)** Slope; **(B)** aspect; **(C)** altitude; **(D)** relief.

## 1.2 Comparison between the discovered terrain-precipitation relationship and baselines

In this work, a spatially varying formula, Eq. (1), is discovered to describe the terrain-precipitation relationship. To better demonstrate the superiority of the discovered formula, we compare it to some baselines and previously proposed empirical terrain-precipitation formulas. The terrain-precipitation formula proposed by Guan et al. is written as follows:

$$P = \omega_1 H + \omega_2 \cos(\beta) + \omega_3 \sin(\beta) + \omega_4 u + \omega_5 v + b, \tag{S.8}$$

where $u$ and $v$ are the longitude and latitude, respectively. As the relief and gradient of altitudes are considered in this work, three baselines are also provided for better

comparison. Baseline 1 is written as follows:

$$P = \omega_1 H + \omega_2 R + \omega_3 \alpha + \omega_4 \beta + b, \tag{S.9}$$

Baseline 2 is written as follows:

$$P = \omega_1 H + \omega_2 R + \omega_3 \alpha + \omega_4 \beta + \omega_5 \sqrt{H} + \omega_6 \sin(\alpha) + \omega_7 \cos(\alpha) + \omega_8 \sin(\beta) + \omega_9 \cos(\beta) + b, \tag{S.10}$$

Baseline 3 is written as follows:

$$P = \omega_1 H + \omega_2 R + \omega_3 H_x + \omega_4 H_y + b. \tag{S.11}$$

The performance of these equations for fitting the average annual precipitation is provided in Table S1. The empirical equation proposed by Guan et al. can describe the terrain-precipitation relationship to some extent. However, the equation has 6 terms, including latitude and longitude, which diminishes the explanation ability since they are not directly related to the terrain. In comparison, our discovered equation achieves higher accuracy with fewer terms that are directly related to the terrain.

Based on Baseline 1, 2, and 3, the accuracy of multivariate regression without interaction terms is limited. To achieve accuracy that is comparable with ours, more terms are required (Baseline 2), which will weaken the simplicity and interpretability of the equation. Notably, the terms in Baseline 3 are similar to ours, but the accuracy is much lower, which confirms the importance of the interaction terms. Previous works usually seek new variables to enhance the accuracy of the fitted precipitation; however, our work demonstrates that the interaction terms can function well with several fundamental terrain variables. Moreover, taking interaction terms into consideration will significantly expand the optimization space, which means that discovering the optimal equation by human experience and trials is not feasible, while AI-empowered knowledge discovery techniques can handle it well.

**Table S1. The performance of baseline equations for fitting the average annual precipitation.** $N_{term}$ is the number of terms in the equation (including the constant term), $R^2$ is the coefficient of determination, and $RMSE$ is the root mean squared error between the observed and fitted values.

| Equations | $N_{term}$ | $R^2$ | RMSE (mm) |
|---|---|---|---|
| Guan et al. | 6 | 0.9794 | 73.27 |
| Baseline 1 | 5 | 0.9776 | 76.19 |
| Baseline 2 | 10 | 0.9829 | 66.75 |
| Baseline 3 | 5 | 0.9755 | 85.04 |
| Ours | 5 | 0.9823 | 67.92 |

## 1.3 Experiments with different parameters in GA-GWR

In AI knowledge discovery by GA-GWR, there are two important parameters that influence the outcomes, namely, the window length (*wl*) and $l_0$_penalty. The window length controls the influence field of the GWR when calculating fitness, and a smaller *wl* indicates a smaller influence field. In the manuscript, the terrain-precipitation relationship is discovered when *wl*=3, which is a suitable range for studying the influence of mesoscale terrain on precipitation. Here, we also provide the discovered formula with different window lengths, including *wl*=2 and *wl*=4. For *wl*=2, the discovered formula is written as follows:

$$P = \omega_1 H + \omega_2 H_x \cos(\alpha) + \omega_3 H_y \cos(\alpha) + b . \tag{S.12}$$

For *wl*=4, the discovered formula is written as follows:

$$P = \omega_1 \sqrt{H} R + \omega_2 R + \omega_3 H_x + \omega_4 H_y + b . \tag{S.13}$$

We observed that the discovered equation may vary slightly depending on the window length used for analysis. However, despite these minor variations, the main influencing terms remain consistent across different window lengths. Specifically, when *wl*=2, only nearby points participate in the regression, resulting in the equation being too local and simple and lacking generalization ability. When *wl*=4, the impact of further points is accounted for, but this weakens the impact of local terrain. Therefore, *wl*=3 is more appropriate for the discovery of the terrain-precipitation relationship.

Another important parameter is $l_0$_penalty, which controls the penalty for the number of items in the knowledge discovery. A higher $l_0\_penalty$ usually induces the discovered equation to be more parsimonious with fewer terms. In this work, $l_0\_penalty$ is set as $1 \times 10^{-5}$, which achieves a balance between parsimony and accuracy. Here, we also provide the discovered formula with different $l_0\_penalties$, including $l_0\_penalty=5 \times 10^{-5}$ and $6 \times 10^{-5}$. For $l_0\_penalty=5 \times 10^{-5}$, the discovered formula is written as follows:

$$P = \omega_1 H^{1.5} \cos(\alpha) + b. \qquad (S.14)$$

For $l_0\_penalty=6 \times 10^{-5}$, the discovered formula is written as follows:

$$P = \omega_1 H^2 + \omega_2 H^{1.5} + \omega_3 R\cos^2(\alpha) + \omega_4 H_y + \omega_5 \alpha^2 R + \omega_6 \cos(\alpha) H_x + b. \qquad (S.15)$$

If $l_0\_penalty$ is large (e.g., $l_0\_penalty=5 \times 10^{-5}$), the discovered formula is concise with merely one term, but the accuracy needs to be improved ($R^2=0.9647$). In contrast, if $l_0\_penalty$ is small (e.g., $l_0\_penalty=6 \times 10^{-5}$), there are many terms in the discovered formula, which diminishes the parsimony and interpretability of the discovered equation. Notably, Eq. (S.14) implies that altitude is the most important factor among the terrain parameters that influences precipitation, which is consistent with our understanding.

### 1.4 Experiments with different basic genes

In the GA-GWR, the basic genes define the fundamental variables that form the interaction terms to construct the formula. In this work, we utilize 10 basic genes, including $H$ (0), $\beta$ (1), $R$ (2), $\alpha$ (3), $H_x$ (4), $H_y$ (5), $\sqrt{H}$ (6), $\sin(\alpha)$ (7), $\cos(\alpha)$ (8), $\sin(\beta)$ (9), and $\cos(\beta)$ (10). Notably, the basic genes $H$ and $\sqrt{H}$ are employed to generate terms with decimal order (e.g., $H^{1.5}$). In AI-empowered knowledge discovery, the optimization scope is all potential equations constructed by these basic genes. A noteworthy aspect of our study is that the ultimately discovered equation does not necessarily include all the basic genes. The genetic algorithm automatically selects the most relevant and suitable terms during the evolution process. Here, we provide the discovered equations with different numbers of basic genes for comparison. When there are only 4 basic genes, including $H$ (0), $\beta$ (1), $R$ (2), and $\alpha$ (3), the formula is written as follows:

$$P = \omega_1 H + \omega_2 R + \omega_3 \beta^2 \alpha + \omega_4 \beta \alpha + b. \qquad (S.12)$$

The $R^2$ of this equation is 0.9802. When there are 9 basic genes, including $H$ (0), $\beta$ (1), $R$ (2), $\alpha$ (3), $\sqrt{H}$ (6), $\sin(\alpha)$ (7), $\cos(\alpha)$ (8), $\sin(\beta)$ (9), and $\cos(\beta)$ (10), the formula is written as follows:

$$P = \omega_1 H + \omega_2 \beta^2 \sin(\alpha) + \omega_3 H^{1.5} R \cos(\alpha) + \omega_4 \beta \alpha + b. \qquad (S.13)$$

The $R^2$ of this equation is 0.9803. With the AI knowledge discovery technique, the discovered equation with only 4 basic genes can also describe terrain precipitation; however, the accuracy needs to be improved. Moreover, without gradient terms in basic genes, the accuracy of the discovered equation is limited, which also confirms that the gradient terms $H_x$ and $H_y$ are important for describing the terrain-precipitation relationship.

### 1.5 The influence of climate change on the terrain-precipitation relationship

Through the analysis in the main text, the influence of climate change on the terrain-precipitation relationship is revealed. In our analysis in the main text, we primarily focused on studying the temporal variation in the coefficient $\omega_1(t)$ and discovered the "1995 turning point", which not only reveals that the terrain-precipitation relationship changes as the climate changes but also finds that different terrain regions experienced distinct changes in the relationship between altitude and precipitation in 1995. In this section, we will focus on other terrain-related terms to provide more evidence. Figs. S2−S4 illustrate the variation in the average normalized coefficients $\omega_2(t)$, $\omega_3(t)$, and $\omega_4(t)$ with the end year of the time range for different regions. Similarly, we adopt the coefficient of the normalized terms and take the 30-year average in the region to study the impact of climate change. Detailed information on these characteristic regions is provided below.

The Changbai Mountains (Region 1, 41.375°N to 49.375°N and 121.875°E to 127.375°E), located in North China, are characterized by high elevations and significant variations in terrain. These mountains are known for their critical role in influencing local precipitation patterns, which makes them an ideal region for studying the interplay between terrain and precipitation.

Southeast China (Region 2, 24.875°N to 35.875°N and 113.125°E to 121.625°E), including regions, such as Jiangsu, Zhejiang, Fujian and Guangdong provinces, exhibits a diverse landscape with hills, plains, and coastal areas. Its climate is heavily influenced by the East Asian monsoon, resulting in complex precipitation patterns.

The Indochina Peninsula (Region 3, 9.125°N to 31.875°N and 91.875°E to 107.875°E), encompassing countries, such as Vietnam, Cambodia, and Laos, is a region of extensive lowlands and mountainous areas. Its complex terrain significantly impacts the distribution of rainfall, making it a relevant area to study the influence of terrain on precipitation.

The Indian Peninsula (Region 4, 9.125°N to 25.625°N and 69.125°E to 87.125°E ), comprising India and its surrounding regions, features a diverse geography with plains, plateaus, and the Western Ghats mountain range. The complex terrain influences the Indian monsoon, a critical factor in the region's rainfall variability, making it vital to study the interaction between terrain and precipitation.

The Himalayas (Region 5, 23.875°N to 32.625°N and 71.125°E to 92.875°E), a majestic mountain range, stretches across several countries, including Nepal, Bhutan, and parts of India and China. The unique terrain of the Himalayas significantly influences the South Asian monsoon, shaping precipitation patterns across the subcontinent.

The inland region of Northwest China (Region 6, 36.375°N to 48.875°N and 81.625°E to 106.625°E), which includes areas, such as Xinjiang and Inner Mongolia, exhibits diverse topography, including plateaus and desert regions. This region's complex terrain plays a crucial role in determining precipitation patterns, making it an essential area to study climate dynamics in arid and semiarid regions.

From the figure, it is evident that climate change affects all these coefficients,

which confirms that the terrain-precipitation relationship is closely related to climate change. The relief term is relatively less affected by the "1995 turning points" since the average coefficients' change with years is more complex, with multiple cycles of rise and fall. In contrast, the gradient of altitude-related terms is more affected by the "1995 turning points". In most of the characteristic regions, the trend changed around the mid-1990s.

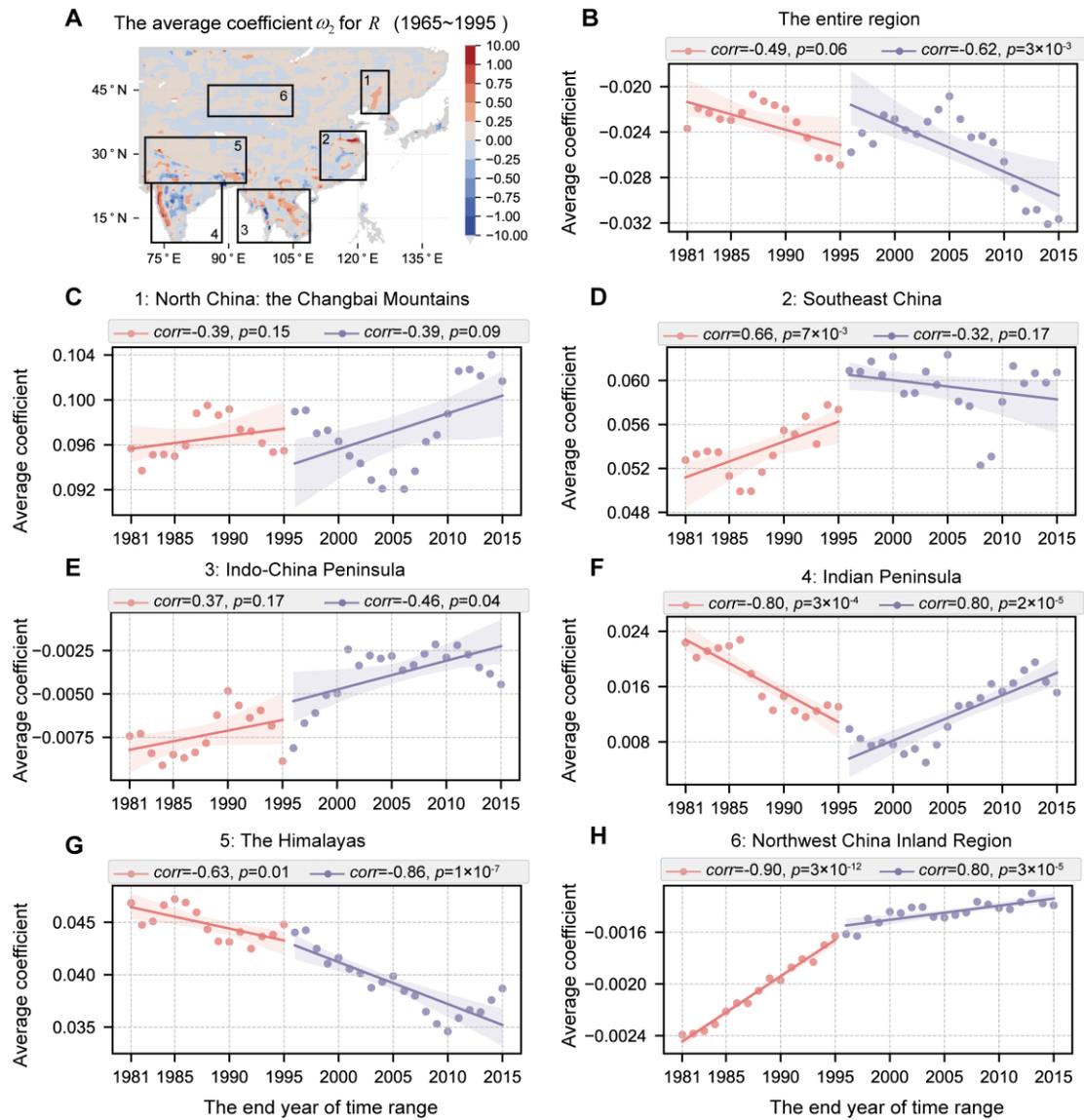

**Fig. S2. The trend of average normalized coefficient $\omega_2$ with years.** **(A)** The investigated characteristic regions. **(B)** The trend of the average normalized coefficient $\omega_2$ in the entire region. **(C)−(H)** The trend of the average normalized coefficient $\omega_2$ in the investigated characteristic regions. The time range is 30 years.

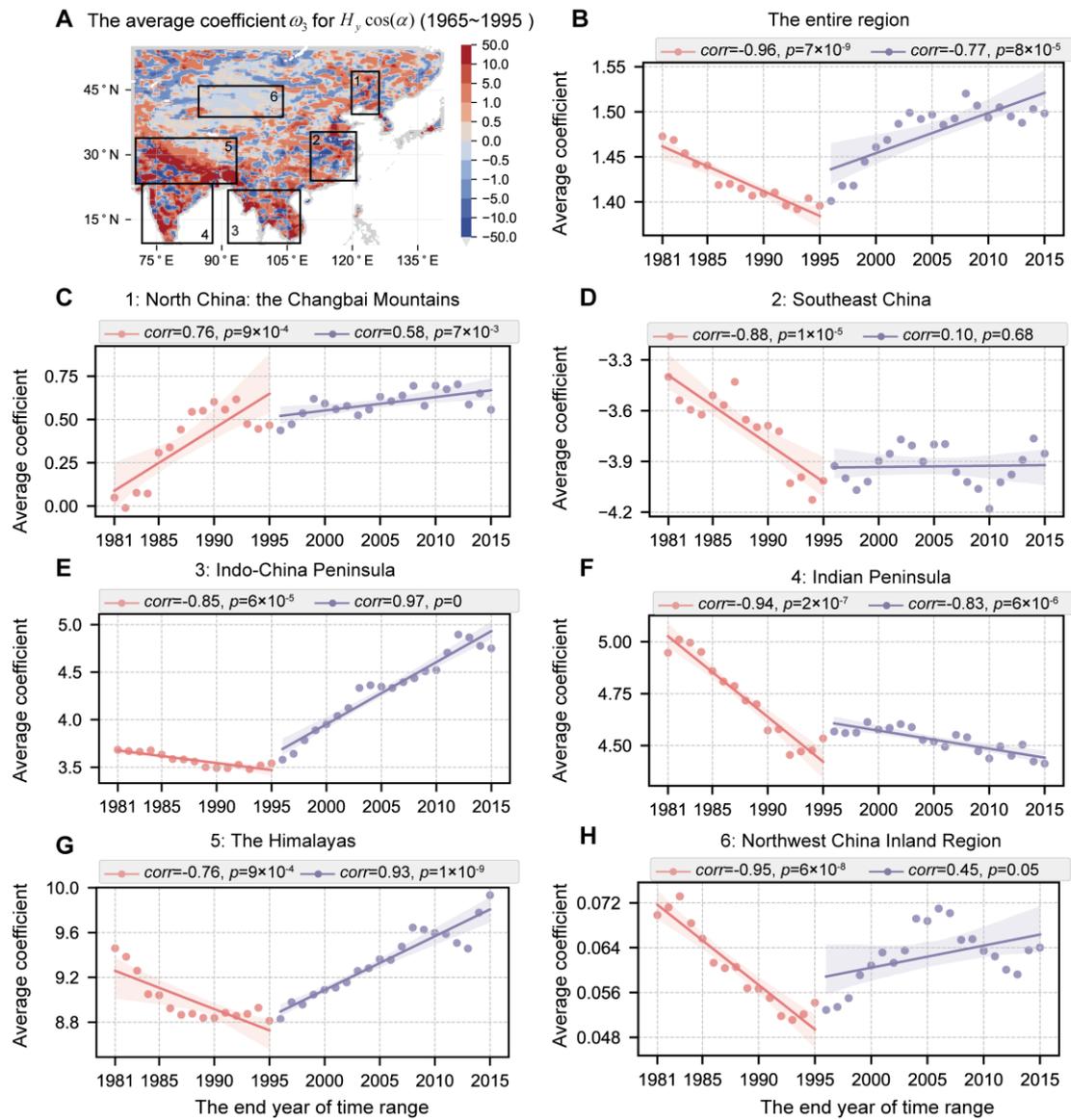

**Fig. S3. The trend of the average normalized coefficient $\omega_3$ with years.** **(A)** The investigated characteristic regions. **(B)** The trend of the average normalized coefficient $\omega_3$ in the entire region. **(C)−(H)** The trend of the average normalized coefficient $\omega_3$ in the investigated characteristic regions. The time range is 30 years.

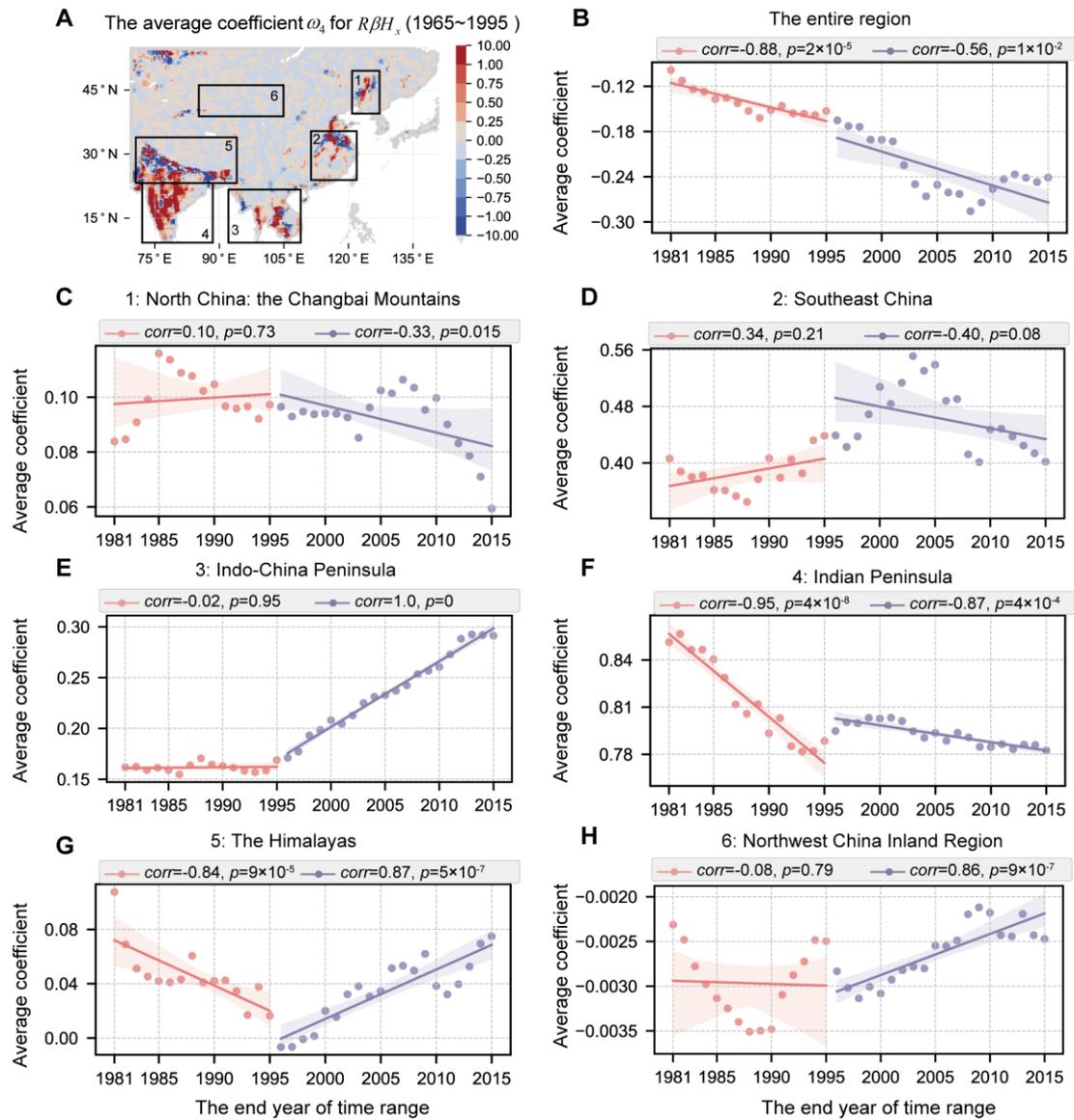

**Fig. S4. The trend of the average normalized coefficient $\omega_4$ with years.** **(A)** The investigated characteristic regions. **(B)** The trend of the average normalized coefficient $\omega_4$ in the entire region. **(C)-(H)** The trend of the average normalized coefficient $\omega_4$ in the investigated characteristic regions. The time range is 30 years.

## 2. Supplementary Figures and Tables.

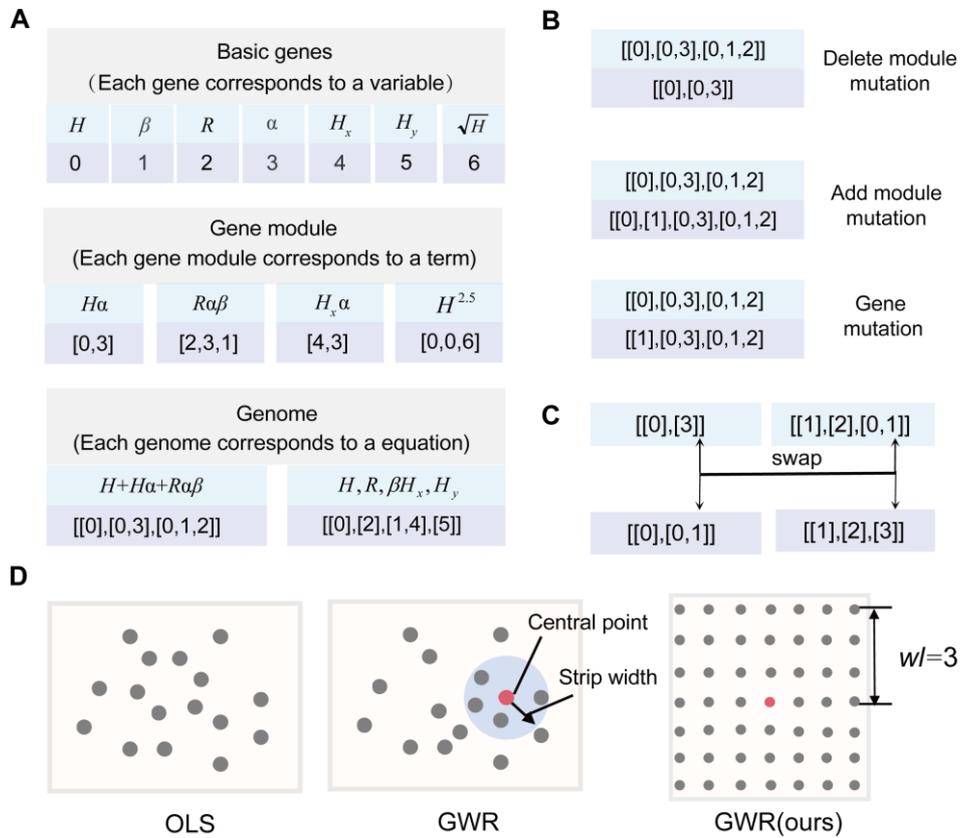

**Fig. S5. Illustration of details in the GA-GWR. (A)** The definition of basic genes, gene module, and genome. **(B)** Examples of the mutation. **(C)** An example of the crossover. **(D)** Illustration of ordinary linear regression (OLS), geographic weighted regression (GWR), and GWR with local regression to estimate the parameters in this work.

**Table S2. Abbreviations of the climate variables in the ERA5 dataset utilized in this work.** The variables in bold are the 23 selected variables.

| | | | |
|---|---|---|---|
| **sp** | **Surface pressure** | swh | Significant height of combined wind waves and swell |
| **t2m** | **2 m temperature** | ssro | Subsurface runoff |
| u10 | 10 m u-component of wind | **sro** | **Surface runoff** |
| u10n | 10 m u-component of neutral wind | tcc | Total cloud cover |
| **u100** | **100 m u-component of wind** | **tciw** | **Total column cloud ice water** |
| v10 | 10 m v-component of wind | **tclw** | **Total column cloud liquid water** |
| v10n | 10 m v-component of neutral wind | lai_hv | Leaf area index, high vegetation |
| **v100** | **100 m v-component of wind** | **lai_lv** | **Leaf area index, low vegetation** |
| **cbh** | **Cloud base height** | **cvl** | **Low vegetation cover** |
| **e** | **Evaporation** | **cvh** | **High vegetation cover** |
| **hcc** | **High cloud cover** | **cape** | **Convective available potential energy** |
| **lcc** | **Low cloud cover** | **mwd** | **Mean wave direction** |
| **i10fg** | **Instantaneous 10 m wind gust** | **mcc** | **Medium cloud cover** |
| **msl** | **Mean sea level pressure** | **pev** | **Potential evaporation** |
| **si10** | **10 m wind speed** | ro | Runoff |
| d2m | 2 m dewpoint temperature | **mwp** | **Mean wave period** |
| **lai** | **Leaf area index** | | |

**Table S3. The evolutionary process when discovering Eq. (1) in the main text.**

| Generation | Discovered terms |
|---|---|
| 10 | $HR, H_x, H_y, \sqrt{H}$ |
| 15 | $HR, H_x, H_y, \sqrt{H}\cos(\alpha)$ |
| 18 | $R, H_x\cos(\alpha), H_y, \sqrt{H}$ |
| 26 | $R, RH_x\cos(\alpha), H_y, \sqrt{H}$ |
| 42 | $H, R, RH_x\cos(\alpha), H_y\cos(\alpha)$ |
| 55 | $H, R, R\beta H_x, H_y\cos(\alpha)$ |
| 100 | $H, R, R\beta H_x, H_y\cos(\alpha)$ |

**Table S4. The evolutionary process when discovering Eq. (2) in the main text.**

| Generation | Discovered terms |
|---|---|
| 1 | $H, R\beta\sin(\alpha), H_x, H_y$ |
| 6 | $H, R, R\beta\sin(\alpha), H_x, H_y$ |
| 11 | $H, R, H_x, H_y\cos(\alpha)$ |
| 24 | $H, R, R\beta H_x, H_y\cos(\alpha)$ |
| 50 | $H, R, R\beta H_x, H_y\cos(\alpha)$ |